\documentclass[intlimits,twoside,a4paper]{article}
\usepackage[cp1251]{inputenc}

\usepackage[eqsecnum]{cmpj3}

\usepackage{graphicx, placeins, bm}

\newcommand{\ga}{\gamma} \newcommand{\be}{\beta} \newcommand{\de}{\delta} 
\newcommand{\la}{\lambda} \newcommand{\al}{\alpha} 
  
\newcommand{\ep}{\epsilon} \newcommand{\s}{\sigma}
  
 \newcommand{\tend}{\rightarrow}

\newcommand{\equa}[1]{\begin{eqnarray} \label{#1}}
\newcommand{\auqe}{\end{eqnarray}}
\newcommand{\auqenn}{\nonumber\end{eqnarray}}
\newcommand{\bloc}[1]{\begin{block}{#1}\begin{center}}
\newcommand{\colb}{\end{center}\end{block}}

\newcommand{\tab}[1]{\begin{tabular}{#1}}
\newcommand{\bat}{\end{tabular} \\ }

\newcommand{\dd}{\hat{m}_i\hat{m}_j -3 (\hat{m}_i \hat{r}_{ij})(\hat{m}_j \hat{r}_{ij})}

\newcommand{\magh}{$\ga$-Fe$_2$O$_3$ }
\issue{2017}{20}{3}{33703}
\doinumber{10.5488/CMP.20.33703}

\title[Ferromagnetic order in dipolar systems with anisotropy]{Ferromagnetic order in dipolar systems with anisotropy: application to magnetic nanoparticle supracrystals\footnote{Paper dedicated to the memory of Dr. J.-P. Badiali.}}
\author[V.~Russier, E.~Ngo]{V.~Russier\thanks{E-mail: russier@icmpe.cnrs.fr}\,, E.~Ngo}  
\address{ICMPE, UMR 7182 CNRS and UPE 2-8 rue Henri Dunant 94320 Thiais, France}

\date{Received June 14, 2017, in final form July 19, 2017, \href{https://doi.org/10.5488/CMP.21.36701}{corrected October 9, 2018}}

\begin{document} 
\maketitle
\begin{abstract}
Single domain magnetic nanoparticles (MNP) interacting through dipolar interactions (DDI)
in addition to the magnetocrystalline energy may present a low temperature ferromagnetic 
(SFM) or spin glass (SSG) phase according to the underlying structure and
the degree of order of the assembly. 
We study, from Monte Carlo simulations in the framework of the effective one-spin or 
macrospin  models,
the case of a monodisperse assembly of single domain MNP fixed on the sites
of a perfect lattice with fcc symmetry and randomly distributed easy axes. We limit ourselves
to the case of a low anisotropy, namely the onset of the disappearance of the dipolar long-range
ferromagnetic (FM) phase obtained in the absence of anisotropy due to the disorder introduced
by the latter.

\pacs 75.50.Tt, 64.60.De
\end{abstract}
\section {Introduction}

The physics of nanoscale magnetic materials is still a very active field of research both 
in view of the potential applications, especially in nanomedicine~\cite{pankhurst_2009}, and  
from the fundamental point of view \cite{skomski_2003,bedanta_2009}. The fundamental aspects are all the more important that 
magnetic nanoparticles (MNP) 
can be synthesized in a wide range of size and shapes and their
assemblies obtained with different structures: colloidal suspensions, or ferrofluids
embedded in non-magnetic materials where one can tune the interparticle interactions through the concentration, or as
powders. Moreover, when the size dispersion is sufficiently narrow, MNP can self-organize in long-range ordered 3D
supracrystals, namely crystals of nanoparticles~\cite{lisiecki_2003b,lisiecki_2011,kasyutich_2010}, 
conversely to the discontinuous metal insulator multilayers (DMIM)
where the underlying structure is disordered~\cite{petracic_2010,bedanta_2013b,chowdhury_2015}.
The interplay between mutual interactions and the degree of order of the underlying structure 
is a key factor in understanding the extrinsic properties of single domain MNP 
assemblies~\cite{bedanta_2009}.
Indeed, due to frustration effects, according to whether the MNP are arranged or not with a long-range order 
at high concentration, a ferromagnetic or a spin glass state is expected at a low 
temperature~\cite{bedanta_2009,petracic_2010} (respectively super-ferromagnetic, SFM and super-spin glass, SSG according to
the nanoscale magnetism nomenclature). Hence, one of the purposes of the modelling is to understand under which conditions the SFM
or the SSG state can be reached.

The modelling of the magnetic properties of both nanostructured materials and materials including nanoscale particles is
a multiscale problem since, on the one hand, the local magnetic structure within MNP
at the atomic site scale presents nontrivial features and, on the other hand, the interactions between MNP play an important role.
An important simplification in the modelling occurs for particles of the radius under a critical size $r_{\text{sd}}$
depending on their chemical composition, typically of a few dozens on nanometers, where they reach a single domain regime 
avoiding both multidomains and vortex structures (typical values for
$r_{\text{sd}}$ are 15~nm for Fe, 35~nm for Co, 30~nm for \magh \cite{skomski_2003}).
In this case, the properties of the MNP ensemble can be described through an effective one-spin model (EOS) 
where each MNP is characterized by its moment and magnetocrystalline anisotropy energy. However, both the moment and
characteristics of the anisotropy term must be considered as effective quantities taking into account some of the
atomic scale features. 
Hence, the effective saturation magnetization $M_{\text s}$ essentially differs from its bulk value due to crystallinity and/or surface defects, the so-called dead layer, and is expected to be smaller than the latter.
As concerns the characteristics of  magnetocrystalline anisotropy energy (MAE), 
the situation is more complex since even its symmetry, either uniaxial or cubic, can differ from 
that of the bulk material. It is worth mentioning that the non-collinearities of the surface spins 
can be represented by an additional cubic term in the MAE in the framework of the EOS~\cite{kachkachi_2006}.

\looseness=-1 The EOS type of approach appears then as a simplifying level of description but, nevertheless, a necessary step 
to model the interacting MNP assemblies, especially when dipolar interaction are included.
In the framework of the EOS models, the theoretical description of MNP assemblies make use of the
standard methods of statistical physics since one deals with a set of particles interacting through 
a pairwise potential in a one-body potential including both the MAE and 
the Zeeman term due to the applied external magnetic field. For MNP coated with a non-magnetic layer,
the interparticle interaction includes an isotropic short-range term and the dipole-dipole interaction (DDI), 
since the exchange term coming from the direct contact between MNP (referred to as superexchange) vanishes.
Hence, one is faced with the models apparented to the dipolar hard or soft spheres, according to the 
short ranged potential (DHS and DSS, respectively) 
widely developed in a more general framework of dipolar fluids including
molecular polar fluids to which one adds the MAE contribution in the form of a one-body potential. 
An important distinction is to be made concerning the structure of the system under study
according to whether it is a liquid-like suspension, a ferrofluid, or a frozen system which can be
an ensemble of MNP in a frozen non-magnetic embedding medium or a powder sample. In this latter  
case, the thermally activated degrees of freedom are the MNP moments and averages over frozen structure
realizations that are done in a second step.

In highly concentrated systems, the nature of the state induced at low temperature 
by the collective behavior resulting from the DDI strongly depends on the underlying 
structure and dimensionality due to the anisotropic character of the DDI.
For a well ordered system such as a fully occupied perfect lattice free of MAE or with a 
MAE characterized by aligned easy axes, a long-range ferromagnetic 
(face centered cubic, fcc, body centered cubic, bcc or body centered tetragonal, bct) 
or anti-ferromagnetic (simple cubic, sc) ordered state is reached 
\cite{luttinger_1946,wei_1992a,weis_1993,bouchaud_1993,weis_2005a,alonso_2010}.
Notice that the solid dipolar ferromagnetic ground state presents a bct structure~\cite{tao_1991a},
a result in agreement with the solid state phase diagram of the dipolar hard sphere system~\cite{levesque_2012a}.
Conversely disordered systems present a spin-glass like (SSG) \cite{bedanta_2013,ayton_1997,petracic_2010}
as has been experimentally evidenced \cite{petracic_2006,parker_2007,wandersman_2008,nakamae_2014,mohapatra_2015}.
It is worth mentioning that not only the perfect lattices of DHS or DSS~\cite{weis_2005a,weis_2006} orient 
spontaneously at low temperature or high pressure. 
This is also the case for the dense liquid state~\cite{wei_1992a,wei_1992b,klapp_2002,klapp_2004,holm_2005}  
where a ferroelectric (or equivalently a FM) nematic state has been reported in the bulk or in slab geometry.

On a perfect lattice, the structural disorder may be introduced either through the MAE
contribution with a random distribution of easy axes or through dilution.
In the limiting case of an infinite magnitude of the MAE, the DHS on a lattice reduces to the
dipolar Ising model, where the exchange coupling constants $J_{ij}$ follow the dipolar energy form
for moments aligned on the easy axes. In the perfect aligned dipole (PAD) case, the ordered/disordered transition takes a SG character when reducing the lattice occupation rate 
\cite{alonso_2010}, while in the random axes dipoles (RAD), the transition is a SG one 
whatever the lattice occupation \cite{fernandez_2008,fernandez_2009}.

Well ordered MNP supracrystals with a fcc structure~\cite{lisiecki_2003b,lisiecki_2011} 
are presumably good candidates for experimental realization of a dipolar induced super-ferromagnetic 
(SFM) phase. The SFM phase was already looked at through the behavior of the hysteresis in terms of temperature
on supracrytals made of iron oxide nanoparticles~\cite{kasyutich_2010}. A ferrimagnetic phase was evidenced involving 
the ferromagnetically ordered MNP core moments and the spin canting at the MNP surface.
In this work we investigate from Monte Carlo simulations the effect of the magnitude of the MAE 
on the dipolar SFM state reached by an ensemble of DHS on a fcc lattice when the MAE easy axes 
are randomly distributed.
We also show that the partial alignment of the easy axes leads to the recovery of 
the SFM state. We only focus on the SPM-SFM transition.

\section {Model and Monte Carlo simulations }

We model an assembly of MNP uncoupled exchange, 
interacting through dipolar interactions (DDI), characterized by their magnetocrystalline 
anisotropy (MAE) and self-organized on long-range ordered supra crystals of fcc symmetry. 
Although this simple model is not at all specific, we have in mind more precisely the case of Co or \magh MNP. 
The MAE is assumed to be of uniaxial symmetry.
This is justified from a crystallographic point of view in the case of hcp-Co, and from the general 
experimental finding for \magh where the shape or the surface effect leads to an effective uniaxial anisotropy.
The easy axes distribution is random unless otherwise mentioned.
In the framework of an effective one-spin model, 
we consider a monodisperse ensemble of up to $N=2048$ 
dipolar hard spheres located on the sites of a fcc lattice
with the volume fraction $\phi$. 
The total energy in a reduced form, after introducing a reference temperature $T_0$ and $\be_0=1/(k_{\text B}T_0)$ reads 
 \equa{ener_1}
  \be_0 E = \ep_d
            \sum_{i < j}  \frac{\dd}{(r_{ij}/d)^{3}}                     
          - \ep_{u} \sum_i (\hat{n}_i\hat{m}_i)^2,               
 \auqe
  where $\hat{n}_i$ and $\hat{m}_i$ are the easy axes and the unit vectors in the direction of the moments, respectively.
  The coupling constants  are related to the saturation magnetization $M_{\text s}$, the uniaxial anisotropy constant $K_u$
  and the MNP volume $v(d)$ [the MNP moment is $\mu$ = $v(d)M_{\text s}$] by
 \equa{coupl_cstes}
 \ep_d=\frac{\be_0 \mu_0}{4\piup} (\piup/6) M_{\text s}^2 v(d) , \qquad \ep_u=\be_0 K_u v(d),
 \auqe
 $\ep_d$ is a usual dipolar constant.
We also introduce more relevant coupling parameters through a reference distance, $d_{\textrm{ref}}=a\phi^{-1/3}$, and a reduced temperature 
\begin{eqnarray} \label{coupl_cstes2}
\lambda_d = \epsilon_d\left( \frac{d}{d_{\textrm{ref}}} \right)^{3}, \qquad \lambda_u = \epsilon_u/\lambda_d\,, 
\qquad T^* = T/(T_0\lambda_d),
\end{eqnarray}
where $\phi$ is the volume fraction and a convenient choice for the proportionality constant is 
$a=(\phi_M^{(0)})^{1/3}$, $\phi_M^{(0)}$ being the maximum value
of $\phi$ for a given structure here taken as the fcc lattice [$\phi_M^{(0)}=\phi_M^{\text{(fcc)}}=\sqrt{2}\piup/6$].
We perform Monte Carlo simulations in the framework of the
parallel tempering scheme~\cite{earl_2005} with a distribution of temperatures
bracketing the superparamagnetic (SPM)-ordered phase transition. 
The set of temperatures $\{T_n\}$ is chosen either from a geometric distribution, 
$T_{n+1}/T_{n}=\text{const}$ or according to the 
constant entropy increase scheme~\cite{sabo_2008} which leads to a more homogeneous rate of 
configuration exchange in the tempering scheme.
Periodic boundary conditions and Ewald summations are used with the so-called conducting external 
conditions, according to which the system is embedded in a medium characterized by an 
infinite permeability, which eliminates the shape dependent depolarizing effects~\cite{allen_1987,weis_1993}.
We emphasize that this is a necessary condition to get a spontaneous magnetization.
At this point, we refer a reader to the analysis and comments in \cite{wei_1992b,klapp_2006}.

We calculate, on the one hand, the mean value of the energy,  the spontaneous magnetization,
per particle 
\equa{m}
\langle m\rangle=\Big \langle \Big| \frac{1}{N}\sum_i{\hat{m}_i} \Big| \Big\rangle,
\auqe
and the nematic order parameter $\la$~\cite{weis_1993}, 
revealing the occurrence of an orientational order. 
The mean value of the magnetization, defined in equation~(\ref{m}) is the order parameter of the
SFM phase. When an anti-ferromagnetic phase is expected, as is the case for the simple cubic (sc)
lattice, the staggered magnetization should be considered in place of $\langle m\rangle$ with
\equa{afm}
\langle m\rangle_{\text{st}}=\left\langle \Bigg\{ 
           \sum_{\al =1,3}  \Bigg[ \frac{1}{N}  \sum_i{\hat{m}_{i\al}(-1)^{p_{i\be}+p_{i\ga}}} \Bigg]^2
           \Bigg\}^{1/2} \right\rangle,
\auqe
where $p_{i\al}$ denotes the layer in the direction $\al$ to which  the lattice site $i$ pertains.

On the other hand, we calculate the heat capacity $C_v$ and the susceptibility from the polarization fluctuation
\equa{dm2}
  \chi_m=N\be\langle \de m^2\rangle =N\be\big(\langle m^2\rangle -\langle m\rangle ^2\big), 
\auqe
both showing a characteristic peak at the transition temperature.
Finally, the location of transition temperature is determined from finite size scaling analysis
through the crossing point of the Binder cumulant~\cite{binder_1997} 
\equa{bm}
B_m = \frac{1}{2}\left(5-3\frac{\langle m^4\rangle }{\langle m^2\rangle ^2}\right),
\auqe
and the location of the $C_v$ and $\langle \de m^2\rangle $ versus temperature peaks.  
Here, as in~\cite{itakura_2003}, $B_m$ corresponds to the normalization such that its limiting values 
in the (anti)-ferromagnetic and paramagnetic phases are $B_m=1.0$ and $B_m=0.0$, respectively. 

Since we consider a random distribution of easy axes, by increasing $\la_u$ we gradually 
increase the amount of disorder. Starting from $\la_u=0$ to $\la_u\gg\la_d$, the model 
goes from the perfect lattice of dipoles to the random axes dipoles (RAD) model where the moments
are assigned in the direction to the easy axes
$\hat{m}_i=s_i\hat{n}_i$ and behave as Ising spins $s_i=\pm 1$. 
The former presents a well ordered ferromagnetic phase at low temperature~\cite{weis_1993}, 
while the latter presents a spin glass state (SSG)~\cite{fernandez_2008}. 
According to this picture, we expect the system to present a SPM~$\tend$~SFM 
and a SPM~$\tend$~SSG transition for small and large values of $\la_u$, respectively.

 \section {Results}
 \indent

In what follows, for the sake of simplicity, we shall denote the SPM and SFM as paramagnetic 
(PM) and ferromagnetic (FM) phases, respectively.
Our purpose is twofold: first, the determination of the PM-ordered phase transition temperature
in the weak $\la_u$ regime, and, secondly, the characterization of the ordered phase. 
In this work we do not focus on the SSG phase which is known to occur at large values 
of $\la_u$~\cite{fernandez_2008}.

In the simulations, we use the parameters of cobalt MNP of ca 8~nm in diameter and the volume
fraction corresponding to the experimental supracrystals elaborated in~\cite{lisiecki_2003b} in
order to fix the dipolar coupling constant. Doing this, we get $\la_d\simeq 2.15$. On the other hand, the
MAE coupling constant is seen as a variable parameter.
It is clear that the conclusions hold for systems characterized by other values of the dipolar
coupling, since we can adjust the value of the reference temperature $T_0$. 
Taking into consideration the fact that the system free of anisotropy presents a FM phase~\cite{bouchaud_1993,weis_1993,weis_2005a},
we expect the ordered phase to keep its FM character at small values of $\la_u$ 
or at least to present the characteristics of a quasi-long-range order (QLRO) magnetic phase 
as it is the case for the random anisotropy model (RAM) in the weak anisotropy regime~\cite{itakura_2003}.

\begin{figure}[!t]
    \begin{center}
    \hskip -0.05\textwidth \includegraphics [width = 0.6\textwidth, angle = -00]{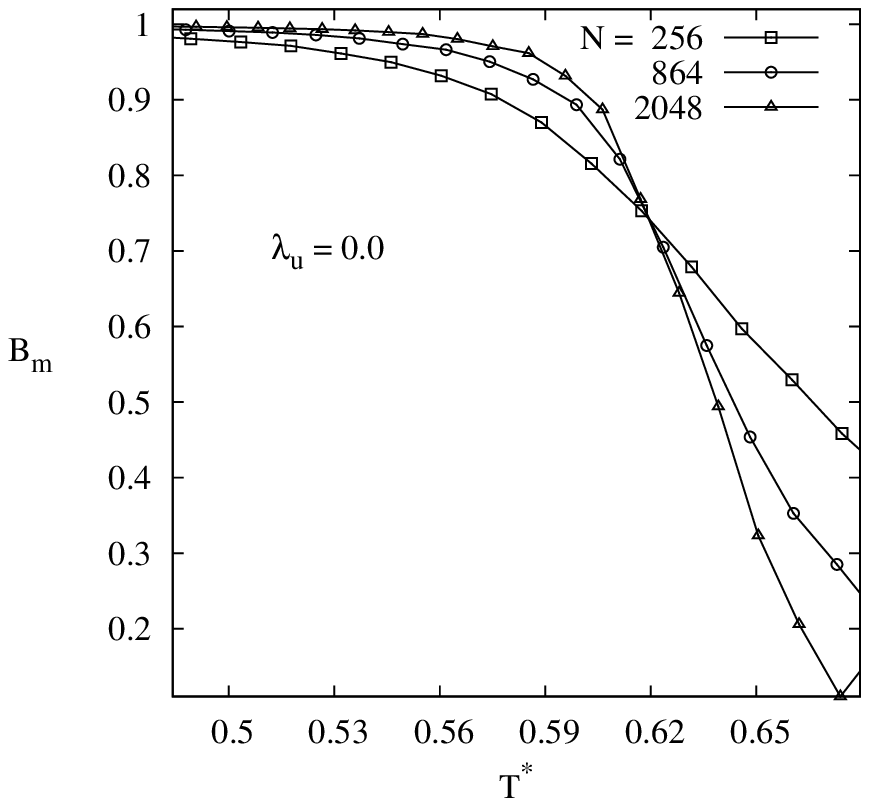}
    \hskip -0.16\textwidth \includegraphics [width = 0.6\textwidth, angle = -00]{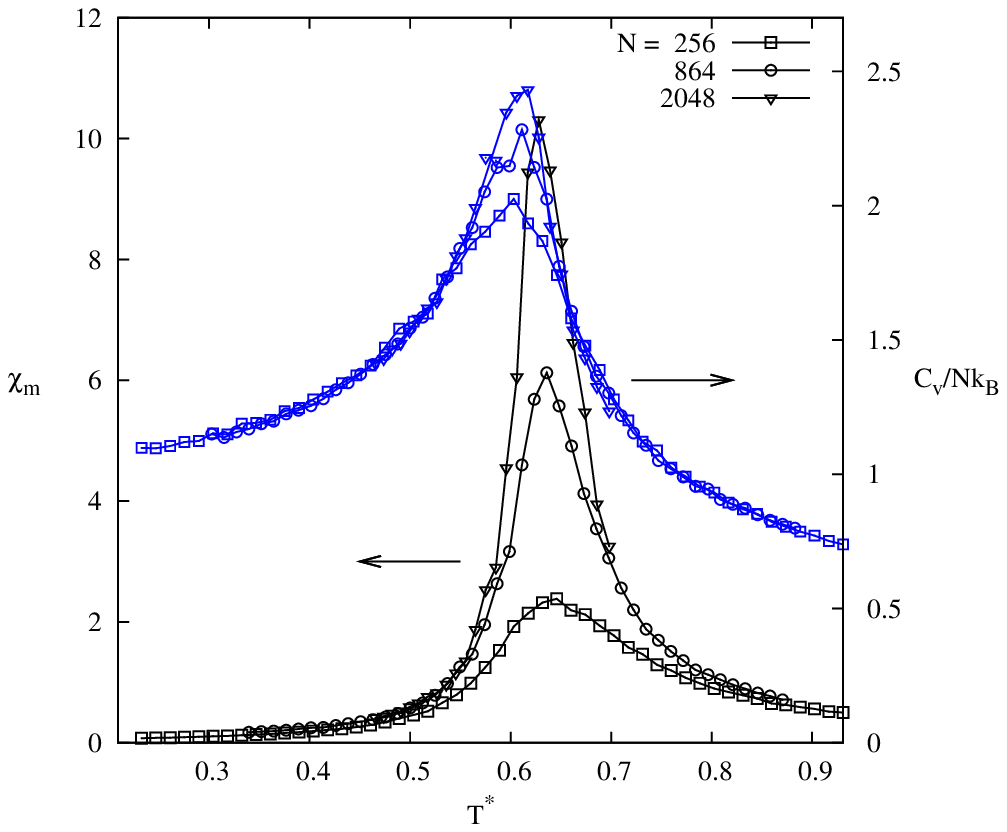}
    \caption{
    \label{binder_lu0} (Color online)
     Left: the Binder cumulant for the moment, equation~(\ref{bm}) for $\la_u=0$.
     Right: susceptibility, $\chi_m$, and  heat capacity $C_v$, right for $\la_u=0$.
    }\end{center}
    \end{figure}

First of all, in the absence of anisotropy, we determine the PM/FM transition temperature in the
framework of the finite size scaling analysis, namely from the crossing point in $T^*$ of the Binder 
cumulant given by equation~(\ref{bm}), 
corresponding to different system sizes, see figure~\ref{binder_lu0}.
     \begin{figure}[!b]
     \vspace{-3mm}
     \begin{center}
\includegraphics [width = 0.6\textwidth, angle = -00]{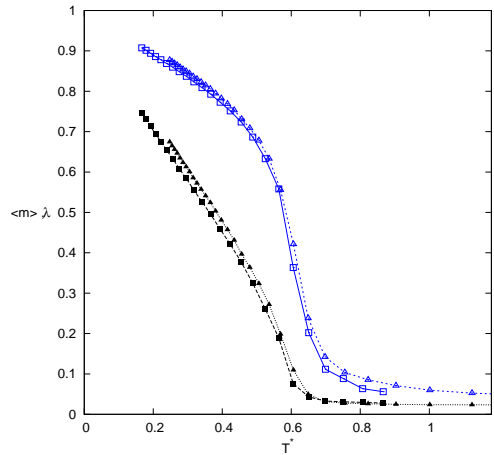}
     \caption{
     \label{pol_cs_cfc}(Color online) 
      Comparison of the polarization (fcc) $\langle m\rangle $ or staggered polarization (sc) $\langle m\rangle _{\text{st}}$,
      open symbols, and nematic order parameter $\la$, solid symbols, on the fcc and sc lattices.
      Triangles: fcc lattice with $N=1372$, squares: sc lattice with $N=1000$,
      $\la_u=0$.
     }\end{center}
     \end{figure}
We locate the PM/FM transition at $T^*_\text{c}=0.618\pm0.005$.
This result, leading to $\rho\mu^2/(k_{\text B}T_{\text c})=2.283$ where $\rho$ is the number density, 
introduced by relating $(d/d_{nn})^3$ of equation~(\ref{coupl_cstes2}) to $\rho$,
is in close agreement with the early finding of Bouchaud and Zerah \cite{bouchaud_1993}.
Moreover, as shown in figure~\ref{binder_lu0}, we find that $T^*_{\text c}$ is very well bracketed by the 
$C_v$ and $\chi_m$ peaks, namely
$T^*_p(C_v)<T^*_{\text c}<T^*_p(\chi_m)$ with $[T^*_p(\chi_m)-T^*_p(C_v)]\simeq 0.02$.
A precise determination of $B_m$ at the PM-ordered transition, namely its non-trivial fixed point value, 
$B_m^*$, is beyond the scope of the present work. It is worth mentioning that this is a difficult task, which is
not unambiguous since it depends on non-essential parameters of the 
transition~\cite{nijmeijer_1996,weis_2006,selke_2007,chen_2004}.
Nevertheless, as expected we get a value, $B_m^*=0.76\pm0.02$ at $\la_u=0$,
compatible with that of either the Heisenberg model on the lattice (0.7916~\cite{holm_1993}) or
the dipolar hard sphere model ($\simeq0.772$ to 0.812~\cite{weis_2006}). 

Then, we compare the behavior of the simple cubic lattice sc to the behavior of the face centered cubic fcc
lattice. It is well known that the sc lattice is antiferromagnetic~\cite{luttinger_1946,fernandez_2000}, 
and thus one should compare the staggered spontaneous polarization, given by 
equation~(\ref{afm}), of the sc to the usual one, equation~(\ref{m}), of the fcc. 
The result for both the polarization and the nematic order parameter $\la$, displayed in 
figure~\ref{pol_cs_cfc} for simulation boxes of similar size shows that the ordering behaviors of the two lattices
present very similar characteristics, at least in the absence of anisotropy. 
This differs from the situation of the Ising PAD model corresponding to $\la_u$~$\tend$~$\infty$ with aligned
easy axes in the $z$ direction where the PM-AFM and PM-FM transitions are found at $T_{\text c}^*\simeq2.5$ and 4.0
for the sc and fcc lattices, respectively~\cite{fernandez_2000}. 

     \begin{figure}[!t]
     \begin{center}
    \includegraphics[width = 0.6\textwidth, angle = -00]{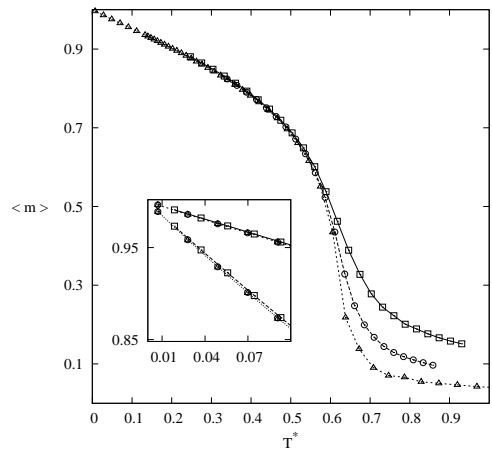}
     \caption{
     \label{pol_eu0}
     Polarization of the fcc lattice in terms of $T^*$ for $\la_u=0$ and
     $N=256$, squares; 864, circles and 2916, triangles.
     Inset: low temperature behavior of the nematic order parameter $\la$, bottom
     and $\langle m\rangle $, top.
     }\end{center}
     \end{figure}
     \begin{figure}[!b]
     \vspace{-3mm}
     \begin{center}
    \includegraphics [width = 0.6\textwidth, angle = -00]{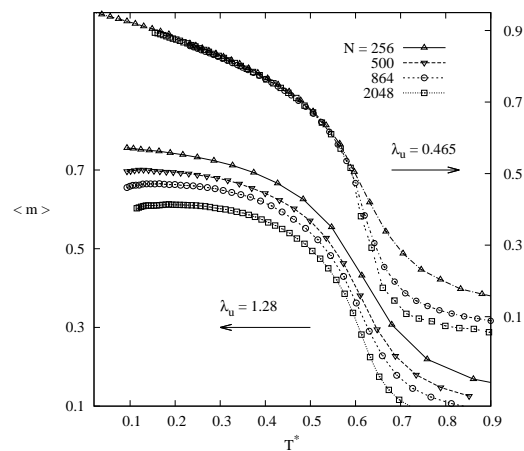}
     \caption{
     \label{pol_lu0.465_1.28}
     Comparison of the fcc lattice polarization in terms of $T^*$ for $\la_u=0.465$ and 1.28
     and different values of $N$ as indicated in the figure.
     }\end{center}
     \end{figure}

Then, we have performed simulations for the fcc lattice and $\la_u$ in the range $0\leqslant\la_u\leqslant1.28$
which is still in the low anisotropy regime. 
We still determine the PM-ordered transition temperature from the finite size behavior of the Binder
cumulant and the location of the $C_v$ and $\chi_m$ peaks. 
An important result is a very weak dependence of the transition temperature with $\la_u$; indeed,
we get $T_{\text c}^*=0.62\pm0.007$ and $0.610\pm0.007$ for $\la_u=0.465$ and 1.28, respectively.
The nature of the ordered phase can be deduced from the low temperature behavior of the 
magnetization $\langle m\rangle $ and the shape of the $C_v(T^*)$ curve. 
We expect the magnetization in the FM phase, on the one hand, to continuously increase 
with a decrease of $T^*$ down to vanishing temperatures and, on the other hand, to be size independent 
below a threshold temperature. 
In particular, the continuous increase of $\langle m\rangle $ down to $T^*\simeq{0}$ is indicative of the absence of 
freezing of the system at any finite temperature.
These two criteria are clearly reached in the absence of anisotropy, as can be seen in figure~\ref{pol_eu0}.
On the other hand, a lambda-like shape of $C_v$ together with the size dependence of the $C_v$ peak,
as displayed in figure~\ref{binder_lu0} for $\la_u=0$ is indicative of the onset of QLRO state from the PM one.
Indeed, the $C_v$ curve shape of typical SG strongly differs from a lambda-like one and, moreover, does not follow
the characteristic finite size scaling of the PM/FM (or PM/AFM) transition~\cite{fernandez_2009,alonso_2010}.
     \begin{figure}[!t]
     \begin{center}
     \hskip -0.05\textwidth \includegraphics [width = 0.59\textwidth, angle = -00]{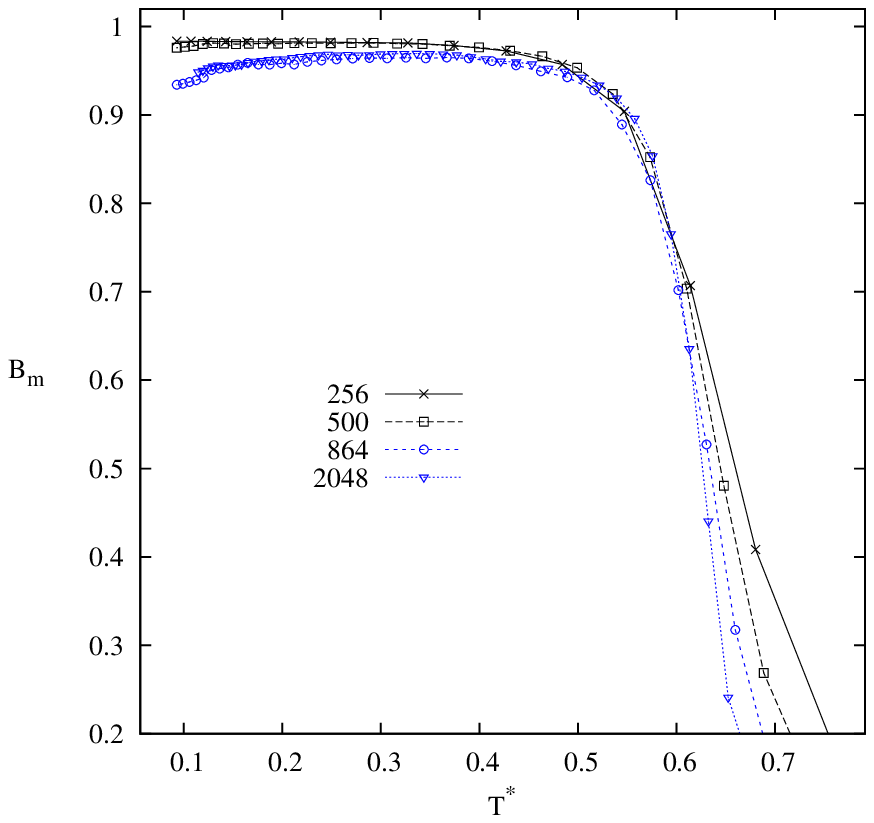}
     \hskip -0.15\textwidth \includegraphics [width = 0.59\textwidth, angle = -00]{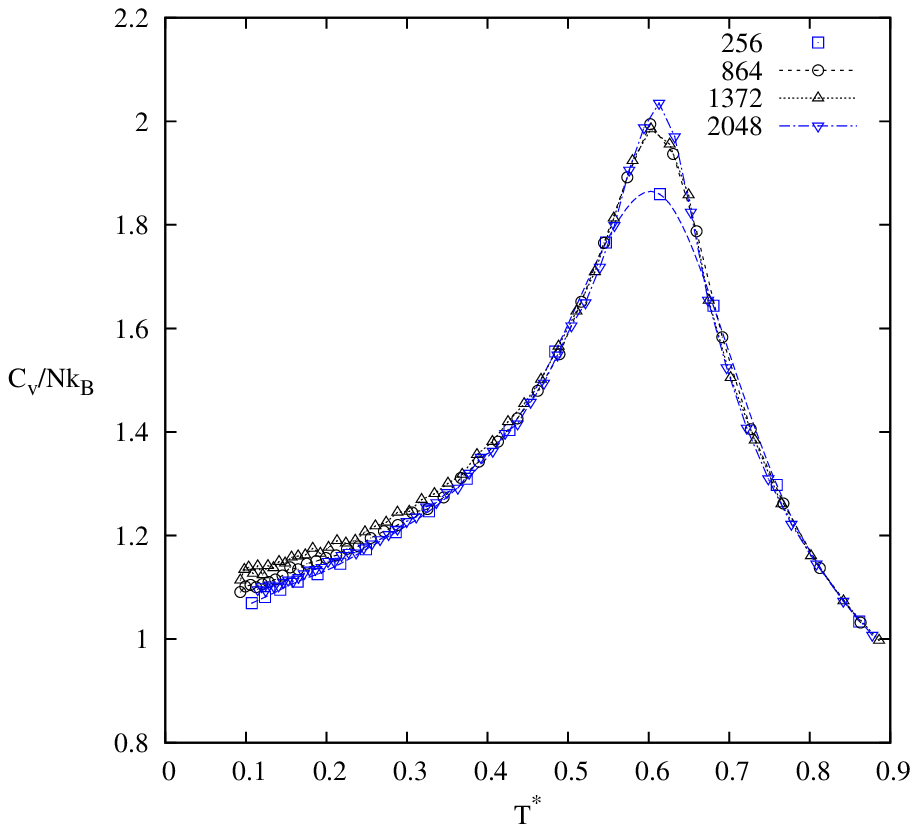}
     \caption{
     \label{binder_ld_2.15_lu_1.28} (Color online) 
     Left:
     Binder cumulant for $\la_u=1.28$ and different values of $N$ as indicated in the figure.
     Right:
     heat capacity for $\la_u=1.28$ and different values of $N$ as indicated in the figure.
     }\end{center}
     \end{figure}
     \begin{figure}[!b]
     \begin{center}
       \includegraphics [width = 0.6\textwidth, angle = -00]{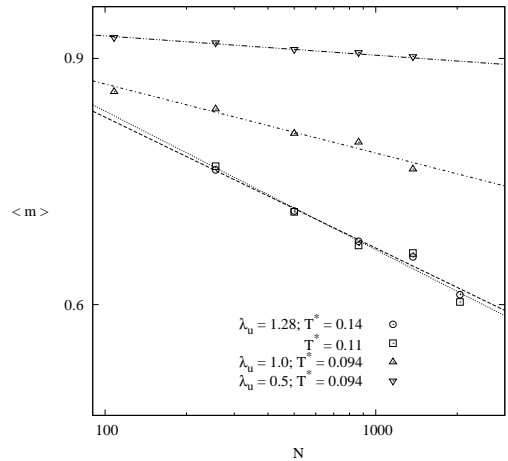}
          \caption{
          \label{pol_log}
          Log log plot of the magnetization at low temperature in terms of $N$.
          The lines are fits of the form $\langle m\rangle =aN^{-b}$ with $b=0.01379\pm0.0009$
          and $0.04936\pm0.006$ at $T^*=0.094$ for $\la_u=0.5$ and 1.0, respectively,
          and $b=0.09355\pm0.0.006$ and $0.09904\pm0.01$
          for $\la_u=1.28$ and $T^*=0.14$ and 0.11, respectively.
          }\end{center}
     \end{figure}
For $\la_u=0.465$, we get a behavior for both the magnetization $\langle m\rangle $ and the nematic order parameter $\la$
very close to that observed in the absence of anisotropy, $\la_u=0$ and the above two criteria are satisfied.
Moreover, the dependence of $\langle m\rangle $ with respect to the number of particles, $N$
is negligible, at least for $N\leqslant2048$, indicating that the mean spontaneous polarization at low temperature 
remains finite on an infinite system. Therefore, we can conclude that the ordered phase is FM in this case.
This is obviously no more the case when $\la_u=1.28$, as shown in figure~\ref{pol_lu0.465_1.28} where
we compare the corresponding spontaneous polarization in terms of $T^*$ for different system sizes. 
Increasing $\la_u$, up to $\la_u=1.28$, the behavior of $\langle m\rangle $ strongly differs and the first point is the 
appearance of a plateau in the $\langle m\rangle (T^*)$ curve at low temperature.
We also get a different behavior for the Binder cumulant, displayed in figure~\ref{binder_ld_2.15_lu_1.28}. Indeed, if there is still a crossing point 
at the onset of the PM-ordered phase, (which is, however, less pronounced),
the low temperature first $B_m$ no more reaches the limiting value $B_m=1$, and its value decreases
with an increase in $N$ which results from the onset of a decrease of $B_m$ with $T^*$ beyond some value of $N$. 
Such a behavior resembles the one obtained in \cite{niidera_2005} where it was analyzed as the occurrence of 
a reentrant SG phase. 
It is worth mentioning that the onset of the PM-ordered transition still presents a FM character, as indicated by
the shape of the $C_v$ curve displayed in figure~\ref{binder_ld_2.15_lu_1.28} but it is presumably a QLRO one instead 
of a FM phase
as indicated first by the size dependence of $C_v$, strongly weakened when compared to the PM-FM
transition case. To confirm this point, we show in figure~\ref{pol_log} the behavior of the $\langle m\rangle $ in terms of $N$ 
at a low temperature. From the fits of $\langle m\rangle (N)$ at fixed $T^*$ in the form $\simeq b.N^{-b}$ and the values of 
$b(\la_u)$ obtained 
($b=0.01379\pm0.0009$, $0.04936\pm0.006$ and $0.09904\pm0.01$ for $\la_u=0.5$, 1.0 and 1.28, respectively), 
see figure~\ref{pol_log}, we conclude that the spontaneous magnetization at a low temperature 
is likely to vanish in the limit $N\tend\infty$ for $\la_u\geqslant 0.5$.
From the low temperature moments configurations obtained with 2048 particles and $\la_u=1.28$, we cannot conclude
on the onset of the domain formation at least in this range of sizes; on the other hand, the $\langle \hat{\mu}_i.\hat{\mu}_j\rangle $
correlation function still presents a FM character with no sign inversion but a noticeably reduced range when compared to 
the $\la_u=0$ case. 

\begin{figure}[!t]
     \begin{center}
     \includegraphics [width = 0.6\textwidth, angle = -00]{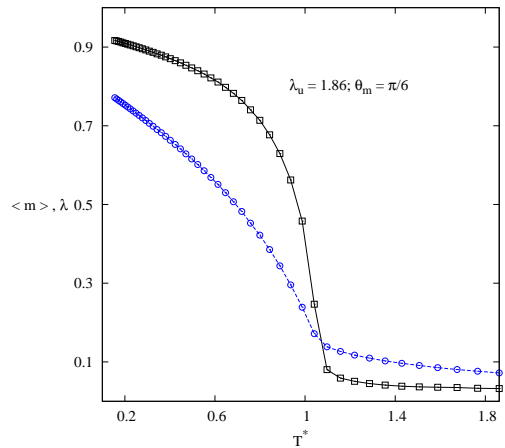}
     \caption{
     \label{pol_text} (Color online)
     Spontaneous polarization (squares) and nematic order parameter (circles) for a system with
     a textured distribution of easy axes centered on the $z$-axis with a solid
     angle limited by a gaussian with variance $\Theta_m=\piup/6$; $\la_u=1.86$.
     }\end{center}
     \end{figure}

Finally, we show that the FM ordered phase is recovered when we consider a preferentially oriented 
easy axes distribution, with probability 
\[ p(\Theta)\propto\sin(\Theta)\re^{-\Theta^2/2\s^2} \] 
around the $z$-axis where one recovers a usual random distribution for $\s\gg1$. In figure~\ref{pol_text} we show an example with $\s=\piup/6$ in the case $\la_u=1.86$ 
a value for which the long-range FM order is lost when the easy axes is randomly distributed.  
%

To conclude, we have shown that an assembly of MNP located on a perfectly ordered fcc lattice presents a 
FM phase at low temperature only in a reduced range of MAE amplitude, $\la_u\leqslant0.5$ for a random distribution of easy axes.
However, there  is an onset of a QLRO phase, at a temperature $T_{\text c}^*$ very close to the PM-FM 
transition of the system free of anisotropy. This QLRO phase is characterized by a spontaneous magnetization
vanishing in the limit of an infinite system. The FM phase can be recovered by aligning, at
least partially, the easy axes.
 
\section* {Acknowledgements}
We acknowledge important discussions with Dr. J.~Richardi,  Dr. I.~Lisiecki, Dr. T.~Ngo,
Dr. C.~Salzemann, Dr.~S. Nakamae, Dr. C.~Raepsaet, Dr. J.-J. Weis and Pr. J.J.~Alonso. 

V.~Russier, who has been one of the Dr. J.-P.~Badiali's PhD students, acknowledges the huge importance
Dr. J.-P. Badiali had on his carrier. The collaboration which lead to about
15 co-authored publications aimed at a theoretical description of the electrochemical
double layer which brought together statistical physics physicists and electrochemists.
It was during this activity that J.P.~Badiali urged V.~Russier to the so-called dipolar hard 
sphere fluid as a model for molecular solvents. The work presented here dealing with another
application of this model is in some sense the continuation of this initial training and collaboration. 

Work performed under grant ANR-CE08-007 from the ANR French Agency.
This work was granted an access to the HPC resources of CINES under the
allocation 2016-A0020906180 made by GENCI. 

\newpage
\ukrainianpart
\title{Феромагнітний порядок у дипольних системах з анізотропією: застосування до магнітних наночастинкових супракристалів}
\author{В. Рус'є, Е. Нго}
\address{Інститут хімії та матеріалів при університеті ``Париж-Схід'' (УПС), дослідницький центр 7182 Національного центру наукових досліджень та УПС, вул. Анрі Дюнана, 2-8, 94320 Тьє, Франція}

\makeukrtitle

\begin{abstract}

Однодоменні магнітні наночастинки (МНЧ) із дипольною міжчастинковою взаємодією можуть окрім магнітокристалічної 
фази знаходитися у низькотемпературній феромагнітній (ФМ) або спіново-скляній фазах залежно від структури та ступеню впорядкування 
системи. Методом моделювання Монте-Карло в рамках односпінової чи макроспінової моделей авторами досліджується монодисперсна система 
однодоменних МНЧ, розташованих на вузлах ідеальної ґратки з симетрією типу fcc і з довільно розподіленими осями. Розглядається випадок 
низької анізотропії, зокрема початку зникнення дипольної далекосяжної ФМ фази.

\keywords магнітні наночастинки, моделювання Монте-Карло, феромагнітне впорядкування

\end{abstract}

\end{document}